\documentclass[lettersize,journal]{IEEEtran}
\usepackage{amsmath,amsfonts,amssymb}
\usepackage{array}
\usepackage[caption=false,font=normalsize,labelfont=sf,textfont=sf]{subfig}
\usepackage{textcomp}
\usepackage{stfloats}
\usepackage{url}
\usepackage{verbatim}
\usepackage{graphicx}
\usepackage{cite}
\usepackage{tikz}
\usetikzlibrary{positioning}

\usepackage[colorlinks=true, allcolors=blue]{hyperref}

\newcommand{\bH}{ {\boldsymbol H} }

\newcommand{\bP}{ {\boldsymbol P} }

\newcommand{\bx}{ {\boldsymbol x} }

\newcommand{\bZ}{ {\boldsymbol Z} }

\newcommand{\bmu}{ {\boldsymbol \mu} }

\newcommand{\bSigma}{ {\boldsymbol \Sigma} }

\newcommand{\bxi}{ {\boldsymbol \xi} }

\begin{document}
	
	\title{Tracking and classifying objects with DAS data along railway}
	
	\author{Simon L. B. Fredriksen$ ^{(1)} $ ,
		T. Tien Mai$ ^{(1),*} $ , Kevin Growe$ ^{(2)} $ ,
		Jo Eidsvik $ ^{(1),*}$
		\\
		{ \small
			$^{(1)}$ Department of Mathematical Sciences,
			Norwegian University of Science and Technology,
			Trondheim 7034, Norway.
			\\
			$^{(2)}$ Department of Electronic Systems,
			Norwegian University of Science and Technology,
			Trondheim 7034, Norway.		
			\\
			$^{*}$Email: the.t.mai@ntnu.no, jo.eidsvik@ntnu.no
		}
	}
	
	\markboth{Journal of \LaTeX\ Class Files,~Vol.~14, No.~8, August~2021}%
	{Shell \MakeLowercase{\textit{et al.}}: A Sample Article Using IEEEtran.cls for IEEE Journals}
	
	
	\maketitle
	
	\begin{abstract}
		Distributed acoustic sensing through fiber-optical cables can contribute to traffic monitoring systems. Using data from a day of field testing on a 50 km long fiber-optic cable along a railroad track in Norway, we detect and track cars and trains along a segment of the fiber-optic cable where the road runs parallel to the railroad tracks. We develop a method for automatic detection of events and then use these in a Kalman filter variant known as joint probabilistic data association for object tracking and classification. Model parameters are specified using in-situ log data along with the fiber-optic signals. Running the algorithm over an entire day, we highlight results of counting cars and trains over time and their estimated velocities.
	\end{abstract}
	
	
	\section{Introduction}
	
	Distributed acoustic sensing (DAS) is a newly developed laser technology where one sends light pulses into fiber optic telecommunication cables, and then measures the backscattered light from impurities in the fiber. Acoustic wave fields generated by for instance a car or train, stretch and compress the fiber cable at event locations. DAS hence transforms a fiber-optic cable into a densely sampled sensor array \cite{culshaw2008fiber}. Even though optical fibers have been in use since the early 1960s, the establishment of Rayleigh-backscattered light for DAS acquisition systems was only popularized within the last decade.  By now, the data are acquired in real-time (at about 2000 times per second) over distances ranging from meters to hundreds of kilometers \cite{waagaard2021rea}. \cite{ip2022using} present a review of recent advances in the field of DAS. 
	
	DAS has seen a diverse set of applications including glacial microseismicity \cite{walter2020distributed}, border control and security applications \cite{juarez2007field}, earthquake detection and characterization \cite{hernandez2022deep} and whale tracking \cite{rorstadbotnen2023simultaneous}. The infrastructure of surplus telecommunication fiber pairs in many cables provides huge potential for various monitoring applications. Already installed fiber-optical cables along roads or railways are naturally suited for traffic monitoring \cite{cedilnik2018advances,milne2020analysis,thomas2023performance,wiesmeyr2020real}. 
	\cite{ferguson2020take} combine positions of DAS events into a prediction system for a train line in Canada, illustrating huge potential for user-oriented products.
	It has however been noted in these studies that extracting DAS measurements along the train track is challenging. This is mainly because of the massive data size ($\sim$ 1 TB of data in a day), and the need for automated signal extraction to reduce this data size. Also, the signals from cars to trains (and other vehicles) vary significantly, and oftentimes there are two or more objects near each other, hampering the allocation of data events to objects. Finally, different types of noise due to, for example, cables installation or fiber impurities are often encountered in DAS data \cite{tribaldos2021surface}. 
	
	In this paper, we develop a multi-object multi-sensor tracking \cite{bar1995multitarget} approach for DAS, assessing car and train traffic along a segment of a fiber-optic cable. 
	Elements required to make the multi-object tracking (MOT) work in real-world settings with massive DAS data include i) Signal processing routines for extracting events from massive-size fiber-optical sensing data. ii) Position and velocity estimation of objects from the DAS events. For step i), we use a combination of frequency filtering, threshold and clustering. We tune parameters of these steps from synchronous data consisting of in-situ logs and DAS data. For step ii), we embed a version of the joint probabilistic data association filter for both tracking and classification of objects. Our suggested workflow enables real-time filtering and prediction of multiple objects moving along the road and reailway near the fiber-optic cable infrastructure. 
	
	In Section \ref{sc_background}, we introduce the necessary background theory about the DAS system, and describe the DAS dataset gathered along the railroad near Trondheim, Norway. 
	In Section \ref{sc_method}, we give a detailed description of the procedure for data signal picking and in Section \ref{jpda} the theory behind multi-object tracking. In Section \ref{sec_class} we extend the MOT approach for classification of events. 
	In Section \ref{sc_results}, we present the results of car and train tracking and classification.

	\section{DAS and Trondheim dataset}
	\label{sc_background}
	
	\subsection{Physical model}
	
	DAS is an emerging sensing technology that employs laser-based mechanisms to detect acoustic waves in real-time along an optical fiber cable \cite{taweesintananon2021distributed}. The cable is transformed into an array of thousands of virtual microphones that monitors acoustic events along the length of the cable almost instantly \cite{liu2017distributed}. 
	Even though it is possible to lay out fiber-optic cables for dedicated purposes of acoustic monitoring, it is most common to use existing tele-communication infrastructure of cable networks. 
	The cable is typically buried underground, but it could also be placed on the surface, or in air or water. 

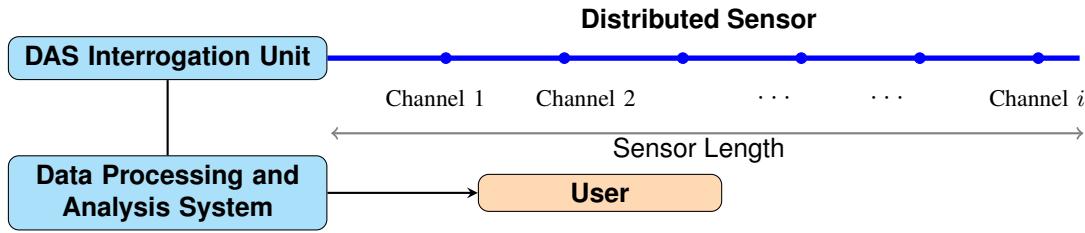
\begin{figure*}[htb]
		\centering
		\begin{tikzpicture}[node distance=2.5cm, font=\sffamily]
		\usetikzlibrary{calc}
		\node (interrogator) [draw, fill=cyan!30, text width=4cm, text centered, rounded corners] {\textbf{DAS Interrogation Unit}};
		
		\draw[thick, line width=2pt, blue] (interrogator.east) -- ++(10,0);
		\node (sensorText) at ($(interrogator.east)!0.5!(interrogator.east -| 12,0)+(0,15pt)$) {\textbf{Distributed Sensor}};
		
		\foreach \i in {0.2,0.4,...,1.2}
		\draw[blue, fill=blue] ($(interrogator.east)!\i!(interrogator.east -| 10,0)$) circle (2pt);
		
		\draw[<->, thick, color=gray] (interrogator.east) ++(0.05,-1) -- ++(10,0);
		\node at ($(interrogator.east)!0.5!(interrogator.east -| 12,0)-(0,35pt)$) {Sensor Length};
		
		\node[font=\small] at ($(interrogator.east)!0.5!(interrogator.east -| 5,0)-(0,15pt)$) {Channel 1};
		
		\node[font=\small] at ($(interrogator.east)!0.5!(interrogator.east -| 9,0)-(0,15pt)$) {Channel 2};
		
		\node at ($(interrogator.east)!0.5!(interrogator.east -| 14,0)-(0,15pt)$) {$\cdot \cdot \cdot$};
		
		\node at ($(interrogator.east)!0.5!(interrogator.east -| 17,0)-(0,15pt)$) {$\cdot \cdot \cdot$};
		
		\node[font=\small] at ($(interrogator.east)!0.5!(interrogator.east -| 21,0)-(0,15pt)$) {Channel $i$};
		\node (processing) [below= 1cm of interrogator, draw, fill=cyan!30, text width=4cm, text centered, rounded corners] {\textbf{Data Processing and Analysis System}};
		
		\draw[thick] (interrogator.south) -- (processing.north);
		
		\node (user) [right=2cm of processing, draw, fill=orange!30, text width=3cm, text centered, rounded corners] {\textbf{User}};
		
		\draw[thick, -stealth] (processing.east) -- (user.west);
		\end{tikzpicture}
		\caption{The DAS system has an interrogator unit (IU) which emits light into the fiber. This returns measurements of the strain at various channels along the fiber over time. Data processing and analysis can turn this acoustic sensing data into a user-friendly decision support system.}
		\label{DASfigure}
	\end{figure*}

 A DAS system consists of an interrogator unit (IU), the distributed sensor (in our case the fiber optic cable) and a data processing and analysis system. The latter enables data processing of strain data, storage, and it provides an interface in which users can interact with the monitoring system. See Figure \ref{DASfigure}.

	The IU emits laser pulses at one end of the cable. The frequency by which the pulses are sent depends on the equipment in place, but the frequency is typically in the thousands of pulses per second. All fiber optic cables contain minuscule imperfections which cause some of the light to be scattered back to the IU, a phenomenon known as Rayleigh backscattering \cite{tribaldos2021surface}. The scattered light forms an interference pattern, which remains constant in the absence of fiber disturbances. In standard optical communication this is considered a nuisance, but in DAS this backscattering is exploited. Acoustic events near the fiber-optic cable cause stretch and contraction in the cable, referred to as fiber strain. This leads to a phase change in the backscattered light detected by the IU. This phase change is measured at equally spaced locations along the fiber, referred to as channels. 
		
	\cite{taweesintananon2021distributed} give a detailed technical description of the DAS system used for gathering data as it was also done in this study. We only provide the core elements here. The light pulses are repeatedly sent with a free-space wavelength of $\lambda_o =1550$ nm, where the sampling period at the optical receiver ($\Delta \tau$) was $10^{-8}$ s. The spatial sampling interval (SSI) is given by the sampling period as
$
	\text{SSI} = \Delta \tau\, c/ (2 n_g),
$
	where $n_g \approx 1.47 $ is the refractive group index of the fiber and $c \approx 3.0 \cdot 10^8$ m$/$s is the speed of light in vacuum. This results in $\text{SSI} \approx 1.02$m, which is the minimum physically possible spatial sampling interval for the system configuration. The measurement however is not a point measurement but a distributed measurement over the so-called gauge length ($L_G$) which describes the length of a fiber section that contributes back-scattered light for a single channel. A larger $L_G$ increases the signal-to-noise ratio (SNR) but reduces the spatial resolution. In addition, to further increase the SNR as well as to reduce the amount of data, channels are commonly averaged and decimated. 
	
	The phase $\phi_x$ of the backscattered light at location $x$ is time-differentiated when obtaining the strain measurements. 
	The measured quantity during the interrogation is the time-differentiated phase change 
   between the two locations $x+L_G/2$ and $x-L_G/2$ and it can be related to axial strain rate:

	\subsection{The Trondheim train track data set}
\begin{figure}[htb]
		\centering
		\includegraphics[width=1\linewidth]{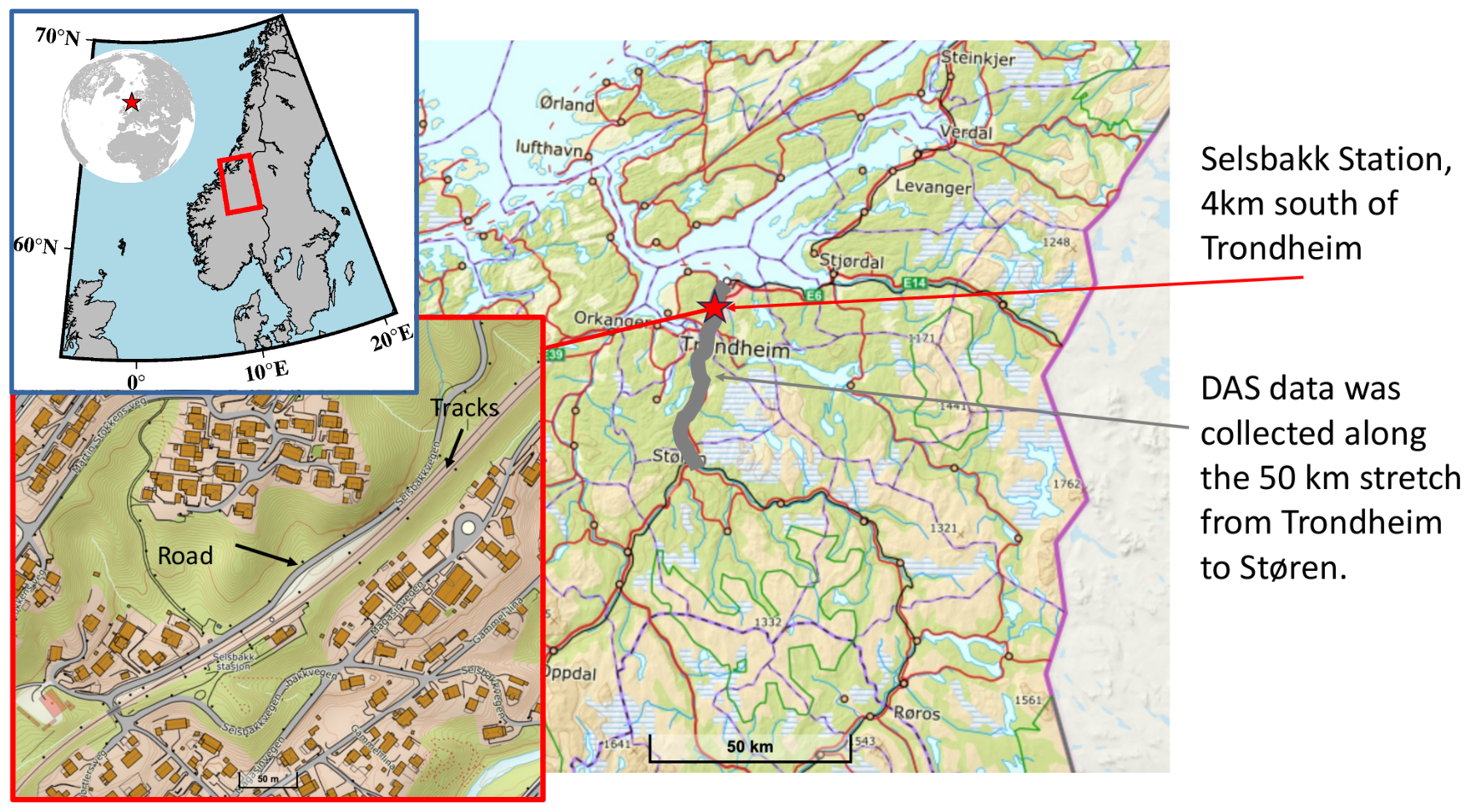}
		\caption{Map of the area around Trondheim, Norway, with the railroad line indicated (grey). The fiber-optic cable is in the ground along the railroad tracks. The main focus is on a segment near Selsbakk station (bottom inset map), where the road runs parallel to the railroad tracks.}
		\label{midnorway}
	\end{figure}

	The DAS data studied in this paper originate from a fiber-optic cable co-located with the 50 km long railway track spanning from Marienborg station to Støren station close to Trondheim, Norway. Figure \ref{midnorway} shows the geographical area of interest. In this work we focus on a 200m long segment near Selsbakk station (inset in display) which is located approximately 4 km south the IU at Marienborg. 
	The data was acquired with an OptoDAS IU \footnote{URL: \url{https://web.asn.com/en/fiber-sensing/main.html}} on the 31st of August 2021. The cable used for the data acquisition is buried in the ground parallel to the railway. Within the train station areas, the cable is encased in concrete, damping its sensitivity. The strain acting on the cables originates from various sources of signals, including those from passing trains, cars, construction, pedestrian traffic and other unknown sources. 
	
	DAS data was acquired with a temporal sampling frequency of 2kHz. 
 As part of the processing, we resampled the data to 1kHz, de-trend it and apply a bandpass filter between 15-150 Hz to filter out mostly lower frequency noise. Instead of using the raw strain data, which contains both positive and negative strain values corresponding to cable compression and extension respectively, we compute its rolling Root-Mean-Square (RMS) in a $0.4$s window length. The RMS is computed on each channel independently, with 50 percent overlap, resulting in one RMS sample every $0.2$s. This data reduction procedure provides a sufficiently accurate description of the signal strength.

	\begin{figure}[htb]
		\centering
		\includegraphics[width=1.0\linewidth]{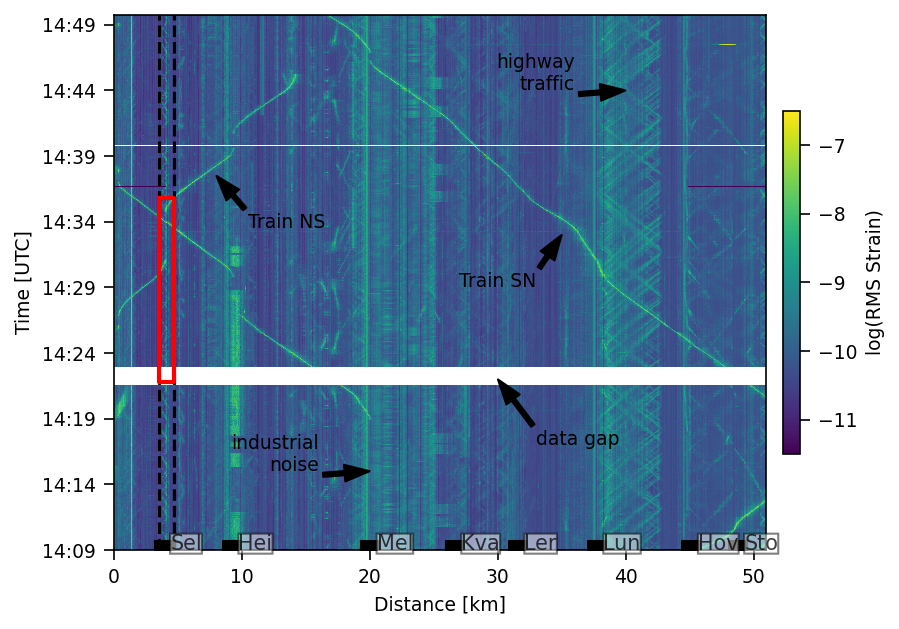}
		\caption{DAS RMS data (log scale) along the entire 50 km (first axis) over a time span of 40 minutes (second axis). Some main events are labeled. NS and SN denote the southbound and northbound trains respectively. The vertical dashed black lines mark the cable section used for this study at a distance at about 4km from the IU. The red frame outlines the data window displayed in Figure \ref{DAS_log_over}. The black squares with labels denote the train stations along the railway.}
		\label{TrondheimDAS}
	\end{figure}
	Figure \ref{TrondheimDAS} shows a heatmap of the DAS log RMS data over the 50 km railway segment.
	In this dataset we notice several events. Trains and cars are recognized by lines of coherent signal strength. Steep line slopes indicate slower velocity. We also notice that acoustic background activity is larger where the fiber runs through urban areas, such as that 10 km from the IU and at 20-25 km.
	
\begin{figure}[htb]
		\centering
		\includegraphics[width=1.0\linewidth]{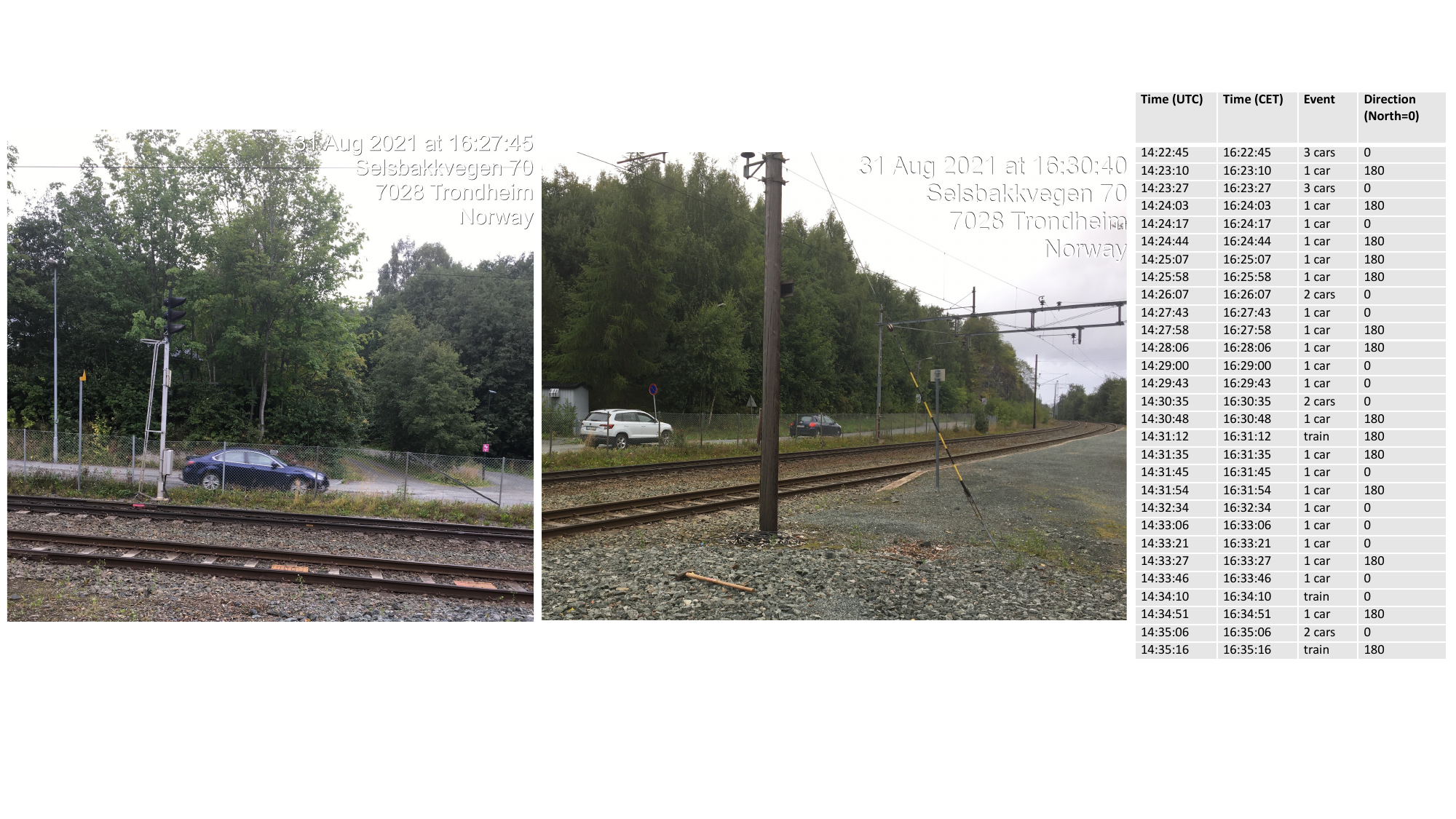}
		 \caption{Events logging at Selsbakk include the times of cars and trains going in different direction. }
		\label{logged_data}
	\end{figure}

	We next zoom in at Selsbakk station 4 km from the IU.
	To ensure labeled data, we logged events over a 13 min time interval on 31 August, 2021. From this work, we made a time-referenced list of cars and trains passing north and south at Selsbakk station (Figure \ref{logged_data}). In the case of multiple cars passing with very short time breaks, the number of cars was also logged with the same timestamp.  	
	
\begin{figure}[htb]
		\centering
		\includegraphics[width=1.0\linewidth]{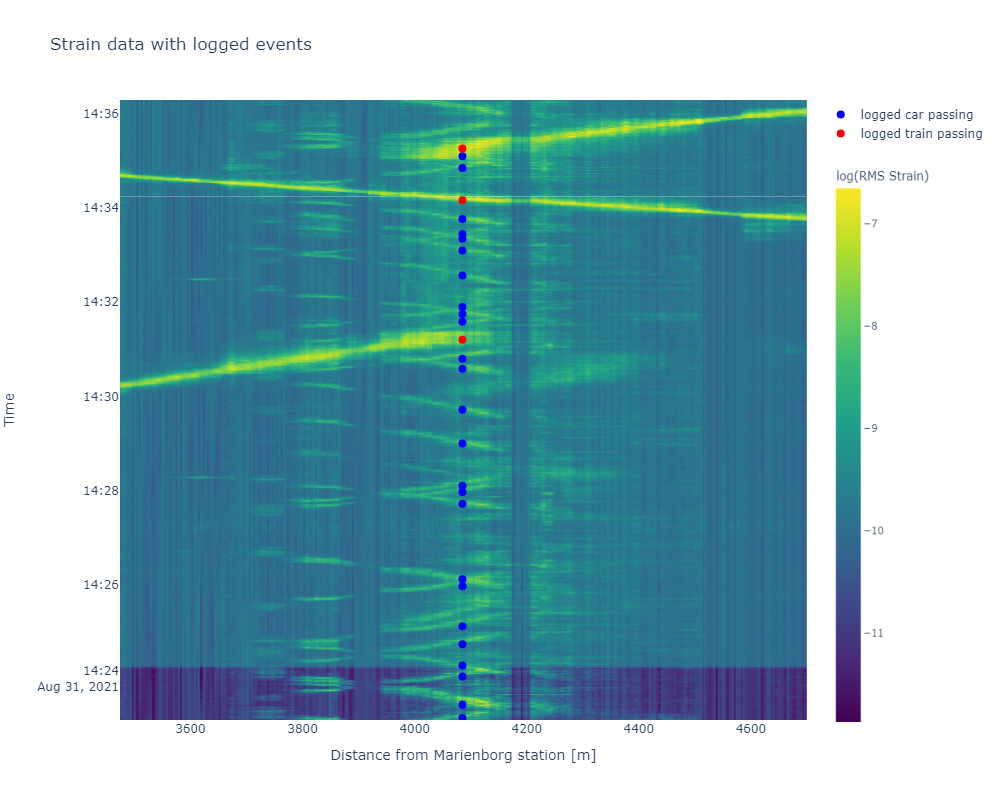}
		\caption{Events logging at Selsbakk together with the DAS RMS data (log scale). This dataset provides labels for the RMS data.}
		\label{DAS_log_over}
	\end{figure}

	In Figure \ref{DAS_log_over} we illustrate these logged events on top of the DAS RMS data at the same time and location. 
	The events consist of three trains (red dots) and about 30 cars (blue dots) driving in the north (0) and south (180) direction. The events appear to align nicely with the increase in DAS data as a consequence of the presence of a vehicle or train. (Note that the ground truth photos in Figure \ref{logged_data} are in local time (CET) which is two hours before UTC time. Later in the article we will stick with UTC time.)
	
	\section{Object detection and signal extraction}
	\label{sc_method}
	
	We now describe the main steps of events extraction. They include i) Pre-processing, smoothing and thresholding, ii) Signal picks clustering, iii) Parameter specification by minimization of the Haussdorf distance between log data and picks.
	
	\subsection{Data processing and detection}
	
	The DAS RMS data are rather noisy because of environmental fluctuations, fiber connection variations and instrumentation errors. Further, acoustic events such as vehicles driving on a road next to a fiber-optic cable generate wave-like disturbances which propagate in the surrounding medium and induce strain variations at multiple consecutive channels and time stamps. Hence, the RMS data do not provide precise information about the source object's exact location at each time. 
	
	In addition to the mentioned frequency filtering and RMS extraction, we apply spatial smoothing over batches of data. The idea is to reduce signal picks that likely originate from the objects. As was done in time, we simply use moving average over a size $\kappa$ window of channels. 
	
	After obtaining a smoothed batch of DAS RMS data, we apply a signal amplitude threshold $A$. This determines a limit for which signals are significantly large to not be considered as noise. 
	Both this threshold parameter $A$ and the window size $\kappa$ are specified by minimizing a distance metric between log data and DAS picks. 	
	Noteably, after the parameters have been pre-trained, the frequency filtering, smoothing and thresholding all take place in batches, that enables almost instant picking that can be analyzed in near real-time. 
	
	\subsection{DBSCAN for extracting picks}
	\label{dbscan}
	
	Even after smoothing and only considering signal peaks with amplitudes above $A$, the data contains plenty picks because of the complex acoustic waveforms caused by a objects. We approach this challenge by applying a 2D clustering algorithm to the (channel,time) picks. 	
	
    Density-Based Spatial Clustering of Applications with Noise \cite{ester1996density}, or DBSCAN, is a popular clustering algorithm. It identifies clusters of high-density regions and classifies potential outliers as noise. The main idea of the algorithm is that for each point in a cluster, the neighborhood of a given radius with the point in its center has to contain a minimum number of points. We define the \textit{$\varepsilon$-neighborhood}, denoted by 
	$ 
	N_{\varepsilon}(p) = \{q \in D | \text{ dist}(p,q) < \varepsilon \}
	,
	$ 
	to be the ball around a point $p$,
	where $D$ denotes the collection of picks, and $\text{dist}(p,q)$ is a distance function. 
	
	We distinguish between two types of points in a cluster $C$: points inside of $C$ (core points) and points on the border of $C$ (border points). We require that for every point $p \in C$ there is a point $q \in C$ so that $p \in N_{\varepsilon}(q)$ and $|N_{\varepsilon}(q)| >= \mbox{MinPts}$ (core point condition), where $\mbox{MinPts}$ is a lower limit for the number of points inside the radius of a point to be considered a core point. Then, point $p$ is \textit{directly density-reachable} from $q$. It is clear that for pairs of core points this condition holds symmetrically, this is however not true if we have one core point and one border point. Furthermore, a point $p$ is \textit{density reachable} from a point $q$ for a given value of $\varepsilon$ and $\mbox{MinPts}$ if there is a sequence of points $q = p_1, p_2, ..., p_n = p$ such that $p_{i+1}$ is directly density-reachable from $p_i$. Also here, the condition is symmetric for pairs of core points, but not generally. 
	In the case where we have two border points in the same cluster, they are possibly not density reachable from each other since the core point condition might not hold for them both. However, there must be a core point from which both of these border points are density reachable, within the same cluster. Using the concept of \textit{density-connectivity}, we ensure that border points in the same clusters are indirectly connected through core points, although they are not directly density-reachable from each other. 
	
	With the concepts of density reachability and connectivity introduced, we can properly define the concept of cluster and noise. A cluster $C$ is a non-empty subset of the collection of points $D$ that satisfies
	\begin{enumerate}
		\item $\forall p,q$ : If $p\in C$ and $q$ is density reachable from $p$ with respect to $\varepsilon$ and $\mbox{MinPts}$, then $q\in C$ (Maximality condition),
		\item $\forall p, q \in C$: $p$ is density-connected to $q$ with respect to $\varepsilon$ and $\mbox{MinPts}$ (Connectivity condition).
	\end{enumerate}
	We define noise as the points in $D$ that do not belong to any of the clusters $C_i$ in $D$, so that
	\begin{equation*}
	noise = \left\{p\in D| \forall i : p\notin C_i\right\}.
	\end{equation*}
	The tuning parameters $\epsilon$ will be specified by minimizing the log data and DAS picks data distance. The hyperparameter $\mbox{MinPts}$ is set to $1$ so that every data point in the clustering scheme can be considered as a separate cluster.

	\subsection{Tuning of data processing parameters by minimizing Hausdorff distance}
	\label{hd_dist}
	
	We evaluate the accuracy of the signal picks from DAS in accordance to the in-situ logged data from Selsbakk station (see Figure \ref{DAS_log_over}). We can view the picked data points and the logged data points as two distinct point sets: DAS picks are denoted $P$ and logged events by $E$. We use a metric than can measure the overall alignment of the points in the two sets. Here, the Hausdorff distance is a good option, as it effectively gauges the dissimilarity between data point sets or curves, see e.g. \cite{rockafellar2009variational}. For our purposes, this works as a loss function when training the hyperparameter selection used to produce the signal picks.
	
    The Hausdorff distance between point sets $P$ and $E$ is defined by 
	\begin{eqnarray}
	\label{hddist}
	d_H(P,E) := \max \left\{\max_{x\in P} d(x,E), \max_{y\in E} d(y, P)\right\}, 
	\end{eqnarray}
	where $d(a, E) := \min_{b\in E} d(a,b), a \in P $ and $ d(c, P) := \min_{b\in P} d(c,b), c \in E $, and $d(a,b)$ is the distance between two points in time.
	Two point sets $P$ and $E$ are close in Hausdorff distance if every point in either set is close to some point in the other set. 
	The application of the Hausdorff distance between the signal picks and logged events at Selsbakk station will give a decent picture of the conformity between signal picks and true event times. 
	
	Note that direct optimization of the distance in \eqref{hddist} will not lead to a satisfactory result for $\varepsilon$ because $|P|$ increases with decreased $\varepsilon$, which in turn will lead to smaller $d_H(P,E)$. Essentially, the more points in $P$, the smaller the maximum distance from a point in $P$ will be to the closest point in $E$. Therefore, we incorporate a penalty term for the number of signal picks to regularize $d_H(P,E)$, as
$
	d_H(P,E) + \xi \left| |P|-|E| \right|
	,
$
	where $\xi > 0 $ was tuned with trial and error. 
	
	Minimizing the Hausdorff distance, the following hyperparameters are set:
	\begin{itemize}
		\item $\kappa$ - window size in moving average smoothing.
		\item $A$ - strain amplitude threshold.
		\item $\varepsilon$ - radius in the DBSCAN algorithm.
	\end{itemize}
	These hyperparameters are trained carefully because we see that their values significantly influence the quality of signal picks. A straightforward grid-based optimization method is employed to specify hyperparameters, yielding $A=-8.8$, $\varepsilon = 0.05$, and $\kappa=30$.
	
	\begin{figure}[htb]
		\centering
		\includegraphics[width=0.9\linewidth]{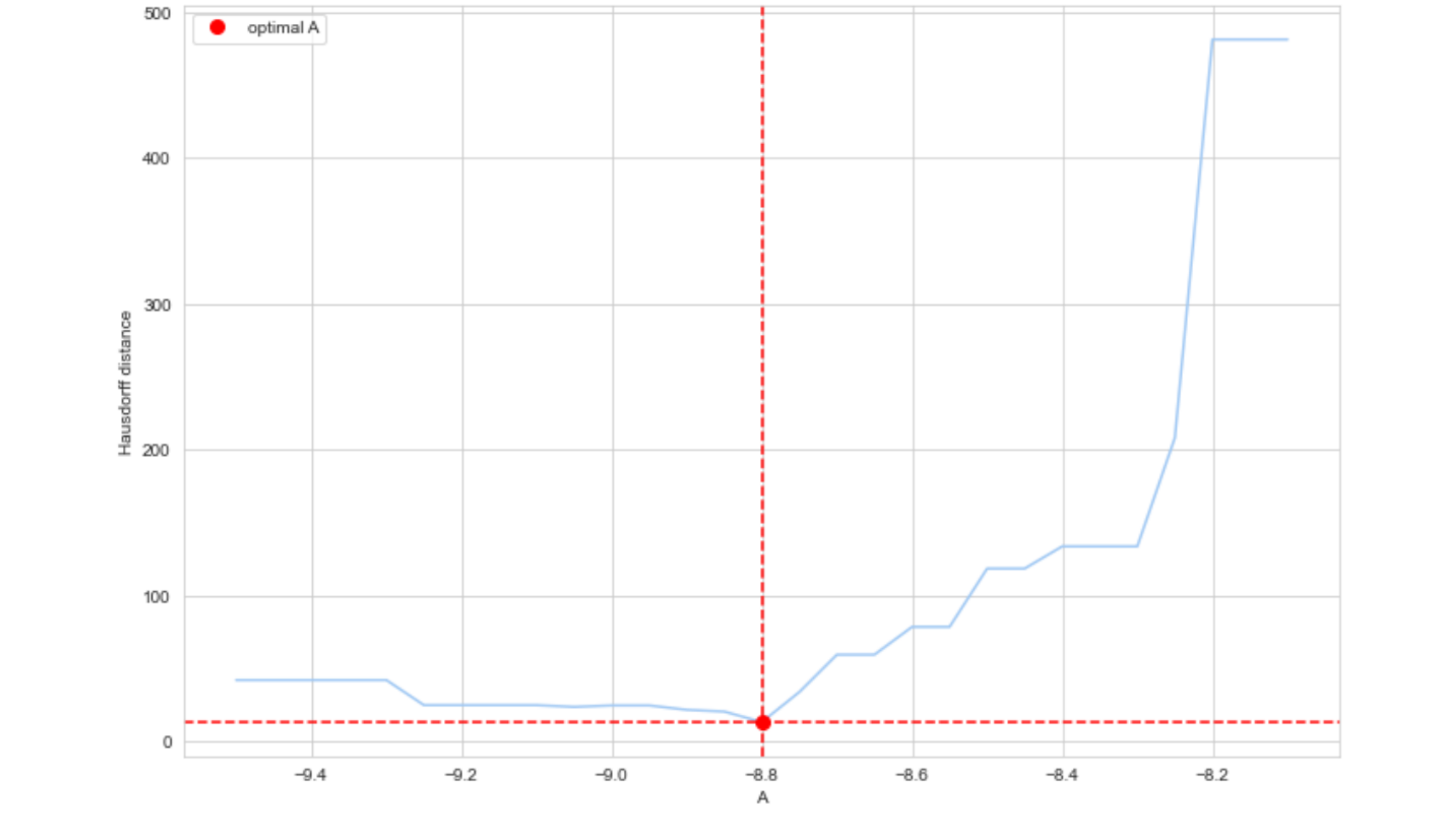}
		\caption{Hausdorff distances for a variety of amplitude threshold parameters.}
		\label{OptA}
	\end{figure}
	
\begin{figure}[htb]
		\centering
		\includegraphics[width=1.0\linewidth]{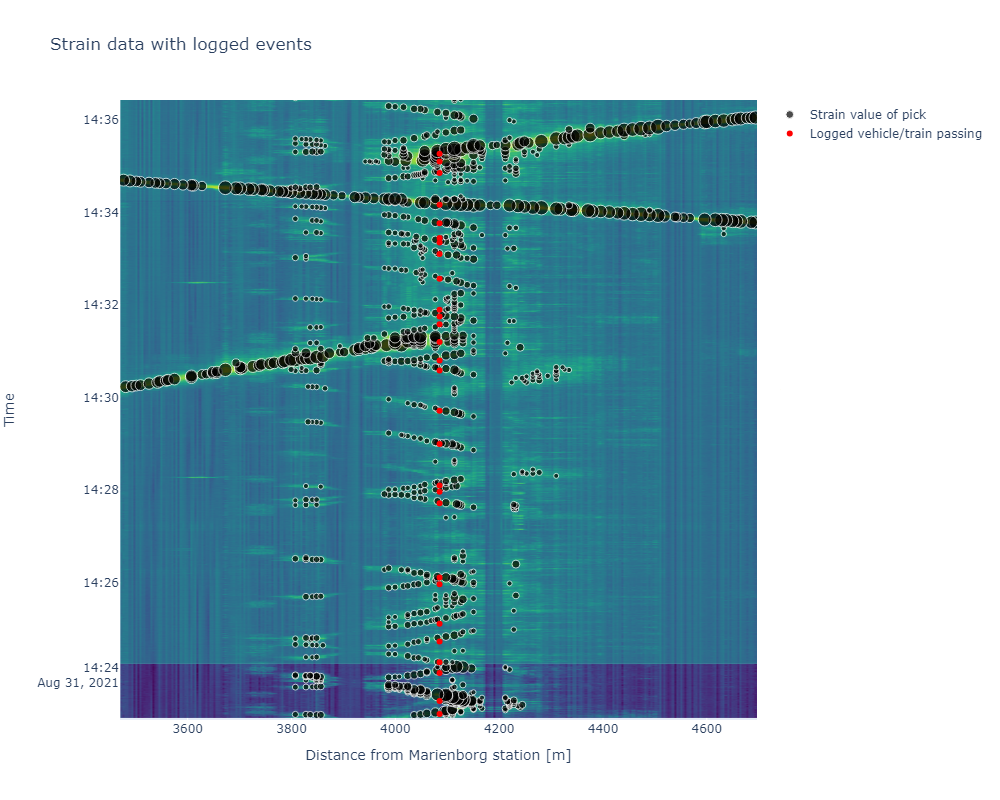}
		\caption{Extracted object events at Selsbakk together with the logged data and the DAS strain data.}
		\label{Heatmap_optconfig}
	\end{figure}

	Figure \ref{OptA} shows the Hausdorff distances for all amplitude thresholds, and we find a minimum at $ A=-8.8$. In this display, $\kappa$ and $\varepsilon$ are fixed to their optimal value.
	Figure \ref{Heatmap_optconfig} shows the resulting signal picks together with the logged events and the RMS signal in the vicinity of Selsbakk station. The picks align quite nicely with the high DAS amplitudes. The amount of data substantially reduced, while keeping the main information in the picks.
    The size of the black circles are proportional to the average strain value in the particular cluster.

	While the hyperparameter values we have determined may result in signal picks that closely align with the logged events on fiber-optic channel 1000 compared, there is no assurance that they will yield the best set of signal picks across all channels. Through further experimentation with hyperparameters and signal picking across all channels, it appears that this configuration may lead to the loss of some important signal picks on certain channels compared to alternative values for $k$, $A$, and $\varepsilon$. In this sense, there could be some undetected cars, but relatively few false positives (detections that are not a real vehicle). Despite this difficult balance, signal picks correlate well with the highest amplitudes on most channels and consequently, we believe that utilizing the optimized hyperparameter values provides sufficient quality when they are utilized as measurements in target tracking over many times and channels.

	\section{Object tracking and the Joint Probabilistic Data Association Filter}
	\label{jpda}
 
	The primary goal of Bayesian filtering is to estimate a latent unobserved state variable from noisy observations \cite{sarkka2013bayesian}. 
	We start by describing the standard situation of one object and one measurement at every time step. We then move to the more complicated situation with multiple objects and measurements, and outline one approximate solution for this setting. Finally, we discuss some choices for implementation.

	A running assumption is that of linear Gaussian dynamic equations for cars and trains, and Gaussian measurement noise for the DAS signal location picks. We denote a Gaussian random variable with mean vector $\boldsymbol{\mu}$ and covariance matrix $\mathbf{\Sigma}$ by $\mathbf{x}\sim \mathcal{N}(\boldsymbol{\mu},\boldsymbol{\Sigma})$, and a Gaussian density function evaluated at vector $\mathbf{x}$ by $\phi(\mathbf{x};\boldsymbol{\mu},\boldsymbol{\Sigma})$.
	
	Multi-object tracking (MOT) entails methods for estimation of position and velocity for many cars and trains in our case, at any time. MOT involves many important concepts and building blocks. Among the more important is data association (DA), which addresses the challenge of allocating measurement picks to objects at every model update time step. In our context, there are challenges with DAS data noise reduction and processing, and careful DA over time becomes important.
	
	\subsection{Single-object tracking}
	
	At time steps $k = 1,2,...,T$, the state $\bx_k=(x_{k,1},x_{k,2})^t$ consists of position and velocity of an object. The goal of tracking is to assess the filtering distribution of $\mathbf{x}_k$ at each time step $k$ based on the history of measurements $\mathbf{Z}_{1:k}=\left\{z_1,...,z_k\right\}$, where $z_k$ is the measurement of position at time $k$. 
	
	The state and observation models are based on the linear Gaussian assumptions, 
	\begin{eqnarray}
	\label{KFssmodel}
	\mathbf{x}_k &=& \mathbf{G} \mathbf{x}_{k-1}+\mathbf{q}_{k-1}, \hspace{1cm} \mathbf{q}_{k-1} \sim \mathcal{N}(\mathbf{0},\mathbf{Q}), \nonumber \\
	z_k &=& \mathbf{H}\mathbf{x}_k + r_{k}, \hspace{1cm} r_{k} \sim \mathcal{N}(0,\sigma^2_r),
	\end{eqnarray}
	with initial conditions $\bx_0 \sim N(\bmu_0,\bSigma_0)$.
	The dynamical model is based on discretization over time with interval step $\Delta_t$. See e.g. \cite{sarkka2013bayesian}. We have
	\begin{equation*}
	\mathbf{G}=\begin{bmatrix} 1 & \Delta t \\ 
	0 & 1
	\end{bmatrix},\hspace{0.5cm} \mathbf{Q}=\begin{bmatrix} \frac{1}{3}\Delta t ^3 & \frac{1}{2}\Delta t^2 \\ 
	\frac{1}{2}\Delta t^2 & \Delta t
	\end{bmatrix}\sigma_q^2, \hspace{0.5cm} \mathbf{H} = \begin{bmatrix}
	1 & 0
	\end{bmatrix}.
	\end{equation*}
	
	The Bayesian filtering distribution has a closed form via the Kalman filter equations. The distribution is Gaussian, and the mean vector and covariance matrix are updated recursively in a one-step prediction and an update step. Assuming the filtering distribution at time $k-1$ is $p(\bx_{k-1}|\bZ_{k-1})=\phi(\bx_{k-1};\bmu_{k-1|k-1},\bP_{k-1|k-1})$, the prediction step at time $k$ is defined by
	\begin{equation}
	\bmu_{k|k-1}= \mathbf{G}\bmu_{k-1|k-1}, \hspace{2mm}
	\bSigma_{k|k-1}= \mathbf{G}\bSigma_{k-1|k-1}\mathbf{G}^T+\mathbf{Q}.
	\label{KFP1}
	\end{equation}
	With the data $z_k$ at time $k$, we have updating (filtering) formula
	\begin{eqnarray}
	\epsilon_k &=& z_k-\mathbf{H}\bmu_{k|k-1}\label{KFU1}, \nonumber \\
	\mathbf{S}_{k}&=&\mathbf{H}\bSigma_{k|k-1}\mathbf{H}^T+\sigma^2_r, \nonumber\\
	\mathbf{K}_k&=&\bSigma_{k|k-1}\mathbf{H}^T(\mathbf{S}_{k})^{-1}, \\
	\bmu_{k|k}&=&\bmu_{k|k-1}+\mathbf{K}_k \epsilon_k, \nonumber\\
	\bSigma_{k|k}&=&\bSigma_{k|k-1}-\mathbf{K}_k\mathbf{H}\bSigma_{k|k-1}, \nonumber 
	\label{Updatestep}
	\end{eqnarray}
	where $\epsilon_k$ is often referred to as the innovation term and $\mathbf{K}_k$ is the Kalman gain. Equation \eqref{KFU1} defines the time-$k$ filtering distribution  $p(\bx_{k}|\bZ_{k})=\phi(\bx_{k};\bmu_{k|k},\bP_{k|k})$, and the recursion continues in real-time over time stages.
	
	\subsection{Multi-object and multi-sensor tracking}
	
	We now go on to describe the situation with multiple objects and multiple data picks, and we are ambiguous about the data-to-object relations. We introduce the various concepts that constitute the MOT framework, see e.g. \cite{vo2015multitarget} or \cite{luo2021multiple} for recent overviews. For simplicity of presentation, we keep the number of objects $n_k$ going into time step $k$ fixed and known.  
	\begin{itemize}
		\item The multi-object states are $\mathbf{X}= (\mathbf{x}_k^1,...,\mathbf{x}_k^{n_k})$. Each target is propagated according to the dynamical model in  \eqref{KFssmodel}.
		\item The observations are $\mathbf{Z}_k=(z_k^1,\ldots,z_k^{m_k})$. They consist of both true object picks and clutter (false positives). The number of observations $m_k$ could be equal to $n_k$, larger then $n_k$ (clutter), or smaller (not detected). 
		\item An object is detected with probability $P_D$, and then the observation model is defined in  \eqref{KFssmodel}. 
		\item Clutter points are distributed according to a Poisson process with constant intensity $\lambda$. 
	\end{itemize}
	
	DA is the mechanism that at any time stage relates an object to a specific measurement. We define DA variable
	$
	h_k=[h_k^1,...,h_k^i,..., h_k^{n_k}]
	$
	where
	\begin{equation*}
	h_k^i = \begin{cases}
	j & \text{ object $i$ is associated to measurement $j$,} 
	\\
	0 & \text{ object $i$ is undetected.}
	\end{cases}
	\end{equation*}
	To ensure consistent DA, a set of rules are defined. Let $\mathcal{H}_k$ be the set of valid DAs at time $k$. For every $h_k \in \mathcal{H}_k$ it must hold that
	\begin{enumerate}
		\item Every object is either detected or undetected: 
$
		h_k^i \in \{0,...,m_k\}, \forall i\in \{1,...,n_k\},
$
		\item Objects cannot share the same measurement:
		$
		\forall i,i' \in \{1,...,n\}, i\neq i', \text{ if } h_k^i\neq 0 \text{ and } h_k^{i'}\neq 0 $, then $ h_k^i \neq h_k^{i'}.
		$
	\end{enumerate}
	
	Putting these DA concepts in the recursive Bayesian filtering scheme, assume that data have been associated with objects up to time $k-1$, with probability $w_{k-1|k-1}^{h}$ for the association hypothesis sequence $h \in \mathcal{H}_{1:k-1}$. This means that the prediction density for the objects is,
	\begin{equation*}
	p(\mathbf{X}_k|\mathbf{Z}_{1:k-1})
	=
	\sum_h w^h_{k-1|k-1}
	\prod_{i=1}^{n_{k}}\phi\left(\mathbf{x}_k^i;\bmu_{k|k-1}^{i,h},\bSigma_{k|k-1}^{i,h}\right),
	\end{equation*}
	where the weights $w^h_{k-1|k-1}$ remain unchanged in the forward propagation, and the mean and variance of each DA hypothesis sequence $h$ follow the usual prediction steps for each object $i$, as explained in \eqref{KFP1}. 
	
	At time $k$, new data $\mathbf{Z}_k$ are available, and DA is used to connect observations and objects. The weights $w^{h'}_{k|k}$ represent the probabilities of a DA sequence $h'=(h,h_k)$ that are now augmented from the one at time $k-1$ ($h$) using the DA at stage $k$ ($h_k$). Note that this hypotheses include the ones where targets are undetected. It also includes possibilities of new objects: hence, and the number of object $n_{k+1}$ that should be propagated grows. Given the hypothesized sequence $h'$ of object-measurement connections, the mean and covariance follow the usual Kalman filter update for each object $i$, so that
	the filtering density is obtained by summing over all hypotheses paths:
	\begin{equation}
	\label{posterior MOT}
	p(\mathbf{X}_k|\mathbf{Z}_{1:k})=\sum_{h'=(h,h_k)} w^{h'}_{k|k}\prod_{i=1}^{n_k} \phi\left(\mathbf{x}_k^i;\bmu_{k|k}^{i,h'},\bSigma_{k|k}^{i,h'}\right).
	\end{equation}	
	The DA problem means that the filtering distribution defined by \eqref{posterior MOT} is a mixture density over a generally very large number of hypotheses. An increase in the number of mixture components occurs at every time step, due to the presence of unknown DAs. Overall, the growing solution turns into a gigantic combinatorial optimization problem \cite{burkard2012assignment} which cannot be handled without using a heuristic approach. There are various MOT methods for handling this issue. We will focus on one solution next.

	\subsection{Joint probabilistic data association}	
	The JPDA algorithm \cite{bar1995multitarget} approximates the expressions in \eqref{posterior MOT} by collapsing the mixture distribution to a single Gaussian distribution. It is based on multiple DA for each target, and then matching the mean and covariance of the resulting mixture distribution at every stage. 
	JPDA hence presents a sophisticated probabilistic approach to the complex problem of MOT. It allows for the simultaneous consideration of multiple potential DA. Unlike other tracking methods that rely on strict associations, like the nearest-neighbor filter, JPDA accounts for the DA uncertainties while still effectively handling the growing complexity. 
	
	A main ingredient of the JPDA approach is the object to measurement probabilities $\beta_k^{i,j}=\text{Pr}[h_k^i=j|\mathbf{Z}_{1:k}]$.
	This probability for each hypothesis includes the outcome of no detection which occurs with probability $1-P_D$. In the case of detecting target $i$, the probability of relating it to $j$ means that the Gaussian observation model is valid for measurement $j$, while the other data are regarded as Poisson distributed clutter for this relation. We then have
	\begin{eqnarray*}
	& \beta_k^{i,0}  \propto  (1-P_D) 
	\\
&	\beta_k^{i,j}  \propto  P_D \frac{\phi(z_k^j;\bH \bmu^i_{k|k-1},\bH \bSigma^i_{k|k-1} \bH^T+ \sigma^2_r)}{\lambda},  j=1,\ldots,m_k.
	\end{eqnarray*}
	The density for object $i$ is determined by using these DA probabilities in the mixture distribution and then computing the mean vector and the covariance. This forms a Gaussian approximation to the filtering distribution. Introducing $\bxi_k^{i,j}=\mathbf{z}_k^j-\mathbf{H}\mu_{k|k-1}^i$, $i=1,\ldots,n_k$, the filtering mean for each target $i$ at time $k$ becomes
 \begin{equation}
     \bmu_{k|k}^i = \bmu_{k|k-1}^i+\mathbf{K}_k^i\sum_{j=1}^{m_k}\beta_k^{i,j}\bxi_k^{i,j},
 \end{equation}
 where the Kalman gain \eqref{KFP1} for this target is $\mathbf{K}_k^i$.
 The variance is
 \begin{eqnarray}
    \bSigma_{k|k}^i &=& \beta_k^{i,0}\bSigma^i_{k|k-1}+(1-\beta_k^{i,0})\bar{\bSigma}_k^i+\Tilde{\bSigma}_k^i \\
    \tilde{\bSigma}_k^i &=& \mathbf{K}_k^i\left(\left(\sum_{j=1}^{m_k}\beta_k^{i,j}\bxi_k^{i,j}\left(\bxi_k^{i,j}\right)^T\right)-\bxi_k\bxi_k^T\right)\left(\mathbf{K}_k^i\right)^T,\nonumber
 \end{eqnarray}
 where $\bxi_k=\sum_{j=1}^{m_k}\beta_k^{i,j}\bxi_k^{i,j}$.
See e.g. \cite{vo2015multitarget} for details around the model updates and other elements related to the JPDA method.

	\subsection{Practical considerations for effective JPDA implementation}
	
	The DA probabilities are in practice only computed within a gate of the predicted locations. This means that the number of entries in the sum is typically smaller than $m_k$, most often only one, two or three.
	
	The field of view (FOV) is restricted to a road/train segment in our case, this provides guidelines for automatic initiation of objects that enter the road segment from one side or the other. The end zones are denoted by lower limit $V_a$ and upper limit $V_b$ along the fiber cable. The initiation is set with a mean $\bmu_0$ at the average location of detections in a boundary zone at the end or start of the FOV, and with the velocity set from observations during in-situ logging. For the covariance matrix $\bSigma_0$, the variance is rather large in both position and velocity, and we use no correlation between the two. 
	
	When it comes to the deletion of tracks, there are various approaches one can take. In this application we have chosen to delete tracks based on the following conditions:
	\begin{itemize}
		\item If  $\boldsymbol{\mu}_{k|k-1}[1] < V_a \text{ or } \boldsymbol{\mu}_{k|k-1}[1] > V_b$, delete the track.
		\item If $\text{trace}(\bSigma_{k|k-1})>\sigma_{\text{thres}}$, delete the track.
	\end{itemize}
	Hence, if the predicted position of the object falls outside the FOV or the predicted covariance of the state exceeds a certain threshold, the object track is deleted. The threshold $\sigma_{thres}$ is chosen by trial and error, and not too small as this will diminish the algorithm's ability to maintain tracks during periods of few updates (low measurement density).

	\section{Classification of objects from amplitudes}
	\label{sec_class}
	
	We now describe a method for not only tracking the location of an object, but also classifying the type of object. In general, similar to the overarching theme of MOT, the type of object, its dynamical model and its observation model can be very complex. One can imagine connecting the object classification to variants of the dynamic Bayesian networks \cite{murphy2002dynamic} or interacting multiple models 
	\cite{mazor1998interacting,johnston2001improvement,challa2011fundamentals}, where the moving target can switch the type of dynamic model over time steps.
	We will follow a simpler approach here, focusing on the measurement amplitudes associated with picks. 
	
	For an object $i$, the class is denoted $\gamma^i \in \{1,2\} $, where $1$ refers to "car" and $2$ refers to "train". When a object is initiated, the prior probability for object $i$ is given by
	\begin{equation*}
	P(\gamma^i = l) = \pi_l^i, \hspace{0.5cm} s.t. \hspace{0.5cm} \sum_{l=1}^2\pi^i_l=1.
	\end{equation*}
	These initial (prior) probabilities of $\gamma_i$ are set using the logged data, i.e. $\pi_1^i=\pi_1$ equals the proportion of cars in the logged data. 
	
	At time $k$, we have $n_k$ objects with unknown class label
	$
	\Gamma = [\gamma^1,...,\gamma^i,...,\gamma^{n_k}].
	$
	The amplitudes associated with the location picks are likely to be informative of the object type. At time $k$, we let the amplitude of the measurement $z_k^j$ be denoted by $y_k^j$. We define a likelihood model for new measurement amplitudes as
$
	p(y_k^j|\gamma^i=l)
	=
	\phi(y_k^j; \alpha_l, \tau_l^2).
$
	Here, the mean values $\alpha_1$ and $\alpha_2$ are specified by the empirical averages over the logged DAS RMS data caused by cars and trains, respectively. Similarly, $\tau_1^2$ and $\tau_2^2$ are specified by the empirical variances of these logged RMS amplitudes.

	\begin{figure}[htb]
		\centering
		\includegraphics[width=1.0\linewidth]{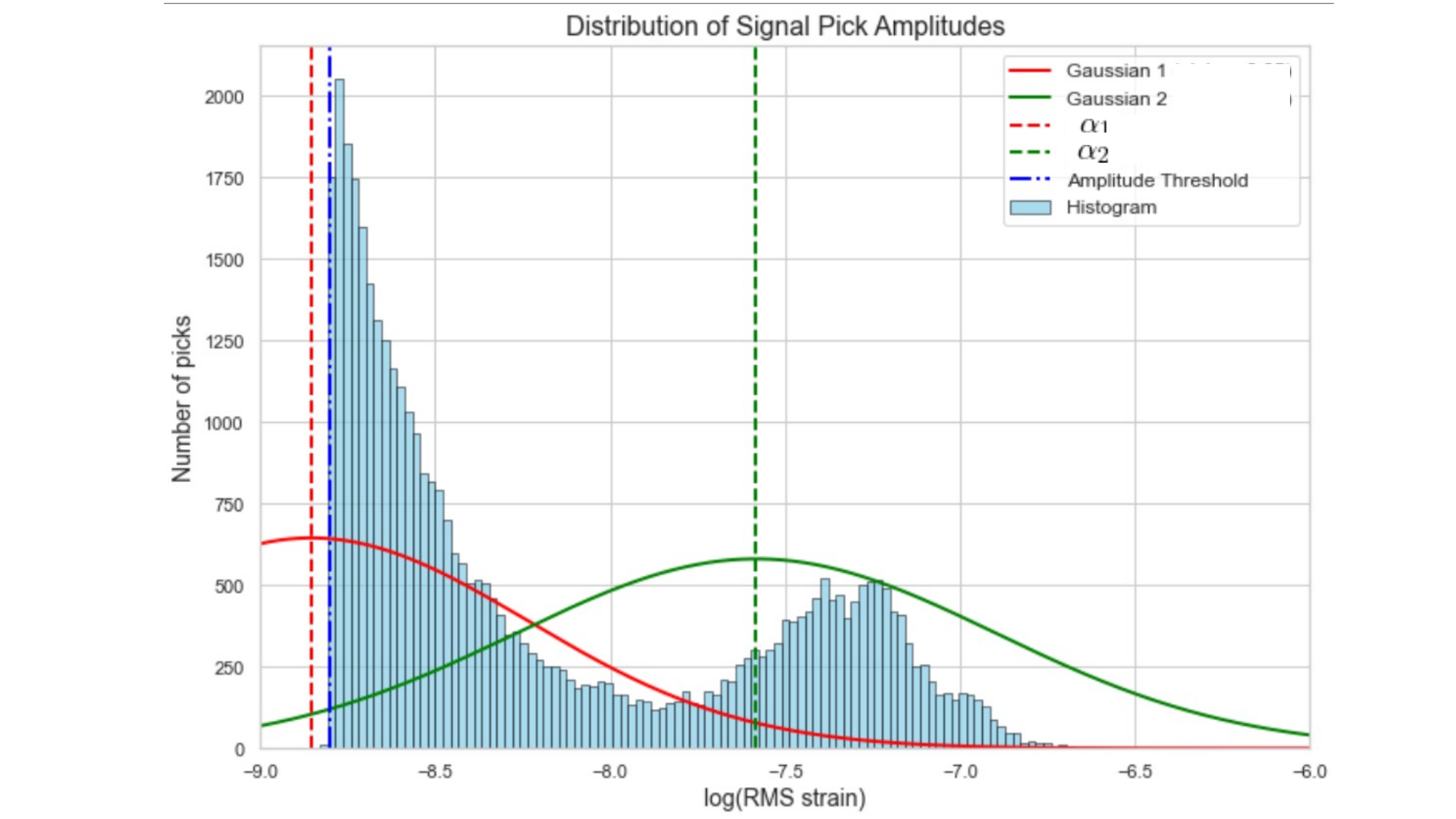}
		\caption{Distribution of signal amplitudes and empirical amplitude densities for cars and trains based on the logged data.}
		\label{amp distr}
	\end{figure}
	Figure \ref{amp distr} shows the histogram of all picked DAS event data along with the fitted (scaled) Gaussian pdf of amplitudes for each class. (The scale is here simply an appropriate factor to make the Gaussian-shaped curve comparable with the histogram of all DAS data picks in the background of the same display.) 
	The empirical data distribution is clearly bimodal. This supports the choice of two classes of objects: cars and trains. Showing all the amplitude data of the processed DAS signals, there are no data below the minimum amplitude threshold $A$ in Figure \ref{amp distr}. 
	The Gaussian distribution for car amplitudes seem to align with the empirical amplitude density for the smallest mode, but we notice that the specified threshold $A=-8.8$ appears to be quite high as it likely remove some of the data for cars. Then again, a lower threshold $A$ would increase the number of false positives, i.e. picks that are clutter and not cars. For trains, we notice that the distribution is slightly shifted to the left compared to the data of train signal picks. There are only three trains in the log, and the estimate of $\alpha_2$ is quite uncertain. Despite slight bias and the difficult balance between thresholding and false positives, the overall shape of the data picks histogram matches the mixture distribution from the log data of trains and cars rather well. 
	
	We now continue to describe the class prediction over time. Let $\mathbf{Y}_{1:k-1}$ denote the sequence of measurement amplitudes up to time $k-1$. 
	The probability that object $i$ is of type $l$, given data up to time $k-1$ is denoted by $\pi_{k-1,l}^i=P(\gamma^i = l | \mathbf{Y}_{1:k-1},\mathbf{Z}_{1:k-1})$, $l=1,2$.
	At time $k$, the data $y_k^j$ are extracted from DAS amplitudes associated with the location measurements $\bZ_k$, and we update the class probabilities for object $i$ by using Bayes' formula:
$
	\pi_{k,l}^i
	=
	P(\gamma^i = l | \mathbf{Y}_{1:k},\mathbf{Z}_{1:k}) 
	\propto
	\left\{ \beta_k^{i,0}+\sum_{j=1}^{m_k}\beta_k^{i,j}\phi(y_k^j;\alpha_l,\tau^2_l) \right\} \pi^i_{k-1,l}
	.
$

	Note that with the bivariate data of both locations and amplitudes; $(z_k^j,y_k^j)$, the DA can rely on both types. The probabilities are then refined to
	\begin{align*}
	\beta_k^{i,j}  \propto 
	 P_D & \frac{\phi(z_k^j;\bH \bmu^i_{k|k-1},\bH \bSigma^i_{k|k-1} \bH^T+ \sigma^2_r)}{\lambda} \times
	\\ 
	& [\pi^i_{k-1,1}\phi(y_k^j;\alpha_1,\tau^2_1)+\pi^i_{k-1,2}\phi(y_k^j;\alpha_2,\tau^2_2)], 
	\end{align*}
	and $\beta_k^{i,0}  \propto  (1-P_D)$.
	In our situation, most ambiguities in DA occur among cars, so the refined probabilities do not matter so much in the practical implementation. 
	
	Recall that one could likely extend this setup and create new transition models based on the class values, but in this paper we assume that both cars and trains follow the same physical dynamics, and therefore have the same transition model. The prediction and update steps in the Kalman filter in Section \ref{sc_method} remain the same, no matter the object type. 
	Possible extensions include different dynamical models for various object types. Say, in a situation where cars move according to this usual model without acceleration terms while trains are expected to accelerate out from a station with an augmented state space model.

	\section{Results}
	\label{sc_results}


	\begin{table}[b]
		\centering
		\caption{Parameter configuration for dynamic model and FOV.}
		\label{groundtruth table DAS}
		\begin{tabular}{|c|c|l|}
			\hline
			\textbf{Notation} & \textbf{Value} & \textbf{Explanation} \\
			\hline
            $\Delta t$ & 0.2 & Time step size. \\
			\hline
			$\sigma_q^2$ & 1 & Process noise parameter. \\
			\hline
			$T$ & 3000 & Number of time steps. \\
			\hline
			$V_a$ & 3963 m& Start of the FOV. \\
			\hline
			$V_b$ & 4167 m & End of the FOV. \\
			\hline
		\end{tabular}
	\end{table}

	\begin{table}
		\centering
		\caption{Parameter configuration for measurement model}
		\label{measmodel table DAS}
		\begin{tabular}{|c|c| p{4cm} |}
			\hline
			\textbf{Notation} & \textbf{Value} & \textbf{Explanation} \\	
			\hline
			$\sigma_r^2$ & 15 & Measurement noise parameter. \\
			\hline
			$P^D$ & 0.9 & Probability of detection for the tracking system. \\
			\hline
			$\lambda$ & $\frac{1}{200}$ & Clutter spatial density calculated from global clutter rate and FOV. \\
			\hline
		\end{tabular}
	\end{table}

	\begin{table}
		\centering
		\caption{Parameter configuration for initiation/deletion of tracks}
		\label{table init delet DAS}
		\begin{tabular}{|c|c|p{4cm}|}
			\hline
			\textbf{Notation} & \textbf{Value} & \textbf{Explanation} \\
			\hline
			$\delta$ & 60m & Initiation range. \\
			\hline
			$\sigma_{\text{thres}}$ & 150 & Covariance threshold for the predicted covariance matrix. \\
			\hline
			$N_{\text{init}}$ & 5 & Minimum number of updates in a holding track for it to be confirmed as a sure track. \\
			\hline
			$\sigma_1$ & 10 & Prior position variance during initiation. \\
			\hline
			$\sigma_2$ & 2 & Prior velocity variance during initiation. \\
			\hline
			$\mu_{v,0}$ & 10 m/s & Prior velocity mean during initiation. \\
			\hline
		\end{tabular}
	\end{table}
	
	We now apply the signal extraction algorithm and the JPDA algorithm to DAS data. The parameters used for the JPDA setup are reported in Table \ref{groundtruth table DAS}, \ref{measmodel table DAS} and \ref{table init delet DAS}. Since the dynamical model of the objects we are tracking can be assumed to maintain a fairly constant velocity, a process noise value of 1 is reasonable. This was also determined through a range of experiments with different values, and we found that $\sigma_q^2$ in the range from 0.25 to 3 gave reasonable results. When it comes to the measurement noise, we expect a high degree of noise, due to the pre-processing, picking and clustering of fiber strain values. We choose $\sigma_r^2=15$ m$^2$, but we found that $\sigma_r^2$ in the range of 5 to 30 gave similar satisfactory results. In our setting with approximate solutions in complex and noisy environments, the choice of these input variance parameters is always a balance, and different parameter values might work better in another tracking scenario. Upon close inspection, we noticed that for the easier setting of night-time with less traffic might benefit from other parameters than the busy day-hours. This suggest an interesting direction for future work in the implementation of non-stationary parameter choices. A position variance of 100 m$^2$ is reasonable given the high noise in measurements, and a lower variance in velocity is reasonable since the objects are expected to move with close to constant velocity. The deletion threshold on the covariance and the minimum number of updates are determined through experimentation. 
	
	Given the parameter choices, we ran JPDA as a filtering exercise on this data. A batch of data was 6 seconds. Starting with the automatic signal extraction and clustering in each batch of data, and then running the JPDA for this time stage, took less than a second. Hence, it is feasible to run the suggested approach operationally in real time (only accounting for the batch size for reliable processing). In practice, of course, communication flow from IU to the analysis center or some kind of edge computing resources is required, but with the effective data reduction to pick the events, we do not consider this to be a bottle neck that would slow down the routines towards getting a real time decision support tool here.
	
\begin{figure}
		\centering
		\includegraphics[width=1\linewidth]{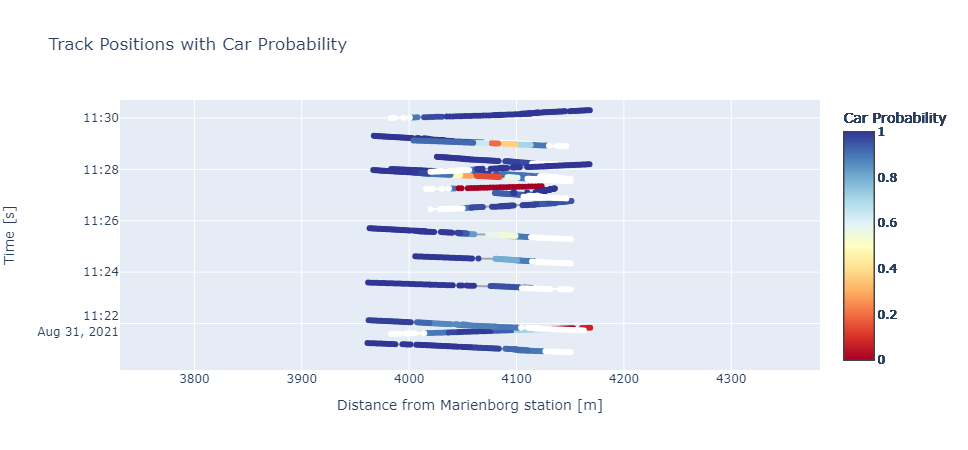}
		\caption{Car probability at each estimated position for all tracks during the time window from 11:20 to 11:30 the 31st of August 2021.}
		\label{idx_class}
	\end{figure}

	Figure \ref{idx_class} shows the tracking results in a selected time window.
	\begin{figure}[htb]
		\centering
		\includegraphics[width=1\linewidth]{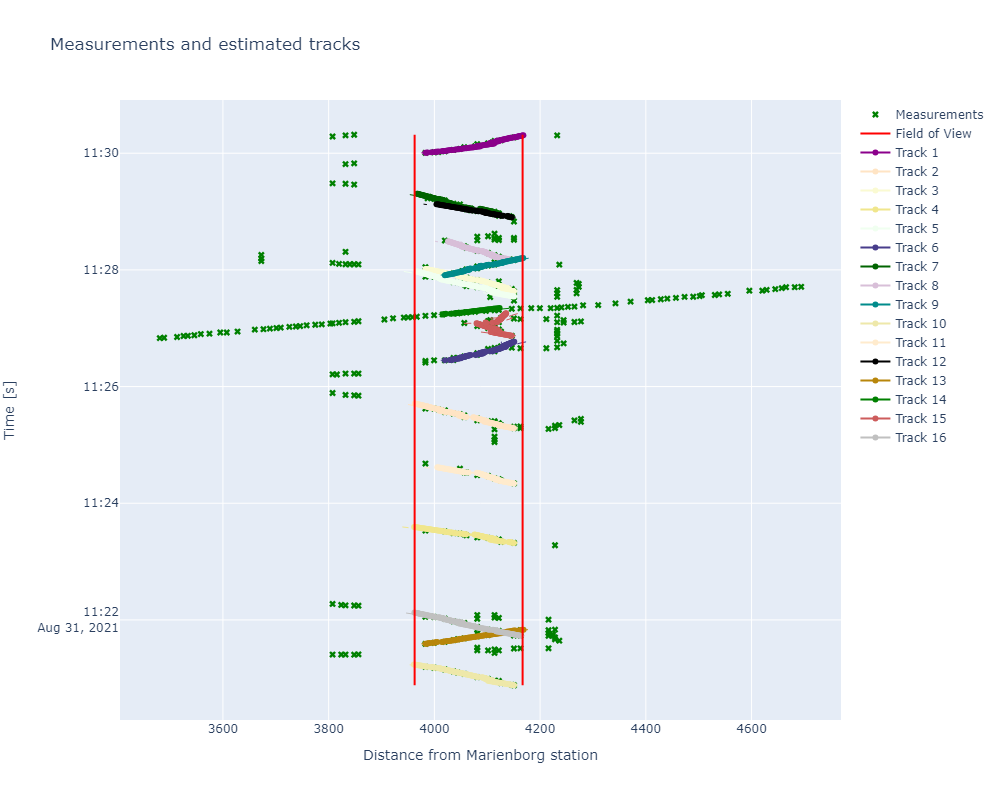}
		\caption{Tracking results of objects in the time window from 11:20 to 11:30 on August 31st, 2021.}
		\label{idx_pos}
	\end{figure}
	The data picks are also shown in this display. The JPDA algorithm is able to track most cars here, but this plot is also showing a problematic case: Around 11:27, when the southbound train enters this area (from the left), there is also a car driving north (red color). In this situation the train will dominate the strain signal on the fiber, and it becomes difficult to detect and track the car successfully. A possibility for improvement here can be to gather more training data and from this attempt to learn the signal characteristics of having car, train, two cars following each other, train and cars simultaneously, etc. Nevertheless, the general pattern from this display and similar plots at other times of the day, indicates that the detection and JPDA algorithm appears to perform very well, especially so when there is limited traffic and vehicles are far apart.
	
	Figure \ref{idx_class} shows results of classification probabilities. Here, high probability of car (blue) and high probability of train (red) get more pronounced as time goes by in the tracking of each object, as is expected because the filtering approach assimilates more amplitude data to confirm the object type.
	Initially, the probabilities can fluctuate, showing the difficulty of classifying some objects as one or the other type. But in most cases, the classification probabilities drift quickly towards $0$ or $1$.

\begin{figure*}[htb]
		\centering
		\includegraphics[width=.7\linewidth]{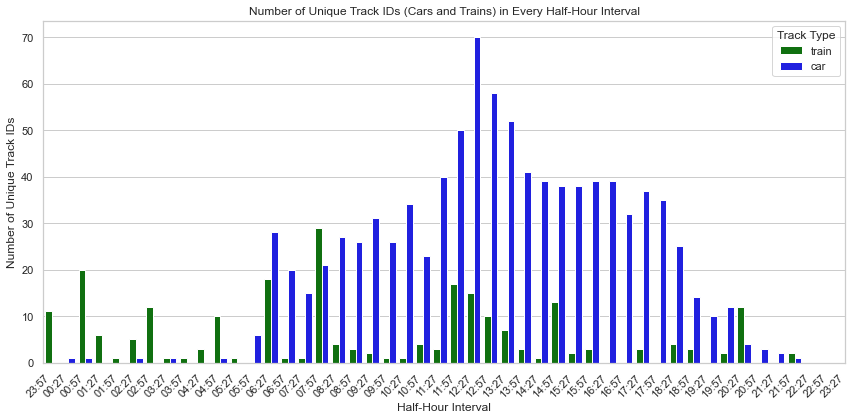}
		\caption{Number of tracked cars and trains during the entire day of 31st of August 2021. Only objects with confirmed track indicator ($N_{init}=5$) are shown.}
		\label{fullday5}
	\end{figure*}

	We next show results of running the detection and object tracking and classification algorithm throughout the data set for the entire day. Figure \ref{fullday5} shows the number (second axis) of tracked cars and trains each half hour during the entire day (first axis).
	As expected the traffic increases significantly from the morning hours and during the daytime, and decreases again around 18:30. The histogram shows that there are many more cars than trains, which is also expected. 

	Traffic seems to peak from 12:30 to 15:00 (14:30-17:00 CET, local time). This is when people drive home from school or work. At this location there is less traffic during the morning. It is worth mentioning that the counts are sensitive to the choice of $N_{init}$ for confirming a track. By visual inspection of tracking resulst within the in-situ logging time period, a reasonable value is $N_{init}=5$. When we decrease this parameter, the counts in all time windows increases, because we allow for shorter tracks to be consider a true track. However, the general traffic pattern throughout the day remains similar for several values of $N_{init}$.
	
\begin{figure}[htb]
		\centering
		\includegraphics[width=1\linewidth]{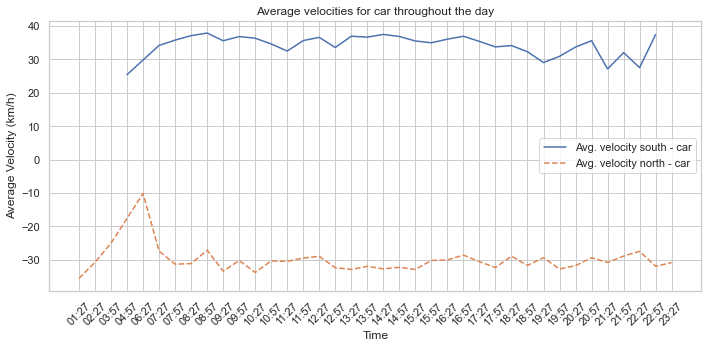}
		\caption{Average velocities of cars driving south and north during the day.}
		\label{avgvel}
	\end{figure}

	Figure \ref{avgvel} shows the average velocities (km/h) of cars driving in both directions. 
	The speed limit is $30$ km/h, but cars driving south enter the FOV right after finishing a $50$ km/h speed limit. This could explain why they are on average driving faster than the one going in the northbound direction.

	\section{Closing remarks}
	
	We have presented a DAS dataset collected on a fiber-optic cable running along the railroad tracks south from Trondheim, Norway. Focusing on a segment of the fiber where the railroad tracks are parallel to the road, we have tracked and classified trains and cars. The results are promising, indicating that DAS could be a highly effective technology for monitoring traffic wherever there is fiber-optic infrastructure. 
	
	The main methodological contributions of this paper involves signal extraction methods using smoothing, thresholding and a DBSCAN algorithm, combined with a JPDA filter for tracking cars. The signal extraction steps ensure that one gets reliable picks; providing a balance of few false positives and not missing true detections. The processing steps hence set the stage for a robust JPDA implementation for object tracking. We further extended this approach to also enable object type classification.  A key benefit of our suggested approach is that it can run in real time. Suitably integrated in a system with the stakeholders, it can then provide an operational decision support tool for traffic monitoring. 
	
	We have shown promising results of tracking and classification of objects. Further, we demonstrated use of this approach for counting cars and trains at different times of day, as well as estimating object velocities. Results indicate that cars coming from the 50 km/h speed limit zone slow down after the 30 km/h sign post, while cars coming from the 30km/h zone speed up at or after this sign for increased speed limit. 
	
	There are challenges with the object tracking method in special situations. 
	Looking at the night hours, there are a number of long trains going by the station. In this case our algorithm that relies on clean picks struggle to track the train. In one example, the JPDA algorithm identifies 13 objects moving in both directions, while in reality this is a long train causing strong fiber strain signals. Parameter tuning could potentially resolve this problem, but a better solution could be to include more flexibility in the measurement model connecting location picks or signal characteristics. On a related matter, some trains also stop at the station, and the JPDA algorithm loose track because the acoustic signal is weak when the train is not moving. A possible solution is the implementation of an switching state model accounting for multiple dynamic behavior regimes over time.

    In monitoring and surveillance, there is a growing need for advanced sensing technologies to provide accurate, real-time data in complex environments, which can enable systems development for operational support. Traditional methods like cameras have limitations such as sparse coverage and GDPR (personal identification) concerns. DAS has evolved into a versatile tool using fiber-optic cables as continuous sensors, and we expect it grow in its use for traffic management as well as in other domains.
	
	\section*{Acknowledgments}
	We acknowledge support from the Centre for Geophysical Forecasting - CGF (grant no. 309960).



	%


\begin{thebibliography}{10}
	\providecommand{\url}[1]{#1}
	\csname url@samestyle\endcsname
	\providecommand{\newblock}{\relax}
	\providecommand{\bibinfo}[2]{#2}
	\providecommand{\BIBentrySTDinterwordspacing}{\spaceskip=0pt\relax}
	\providecommand{\BIBentryALTinterwordstretchfactor}{4}
	\providecommand{\BIBentryALTinterwordspacing}{\spaceskip=\fontdimen2\font plus
		\BIBentryALTinterwordstretchfactor\fontdimen3\font minus
		\fontdimen4\font\relax}
	\providecommand{\BIBforeignlanguage}[2]{{%
			\expandafter\ifx\csname l@#1\endcsname\relax
			\typeout{** WARNING: IEEEtran.bst: No hyphenation pattern has been}%
			\typeout{** loaded for the language `#1'. Using the pattern for}%
			\typeout{** the default language instead.}%
			\else
			\language=\csname l@#1\endcsname
			\fi
			#2}}
	\providecommand{\BIBdecl}{\relax}
	\BIBdecl
	
	\bibitem{culshaw2008fiber}
	B.~Culshaw and A.~Kersey, ``Fiber-optic sensing: A historical perspective,''
	\emph{Journal of lightwave technology}, vol.~26, no.~9, pp. 1064--1078, 2008.
	
	\bibitem{waagaard2021rea}
	O.~H. Waagaard, E.~R{\o}nnekleiv, A.~Haukanes, F.~Stabo-Eeg, D.~Thingb{\o},
	S.~Forbord, S.~E. Aasen, and J.~K. Brenne, ``Real-time low noise distributed
	acoustic sensing in 171 km low loss fiber,'' \emph{OSA continuum}, vol.~4,
	no.~2, pp. 688--701, 2021.
	
	\bibitem{ip2022using}
	E.~Ip, F.~Ravet, H.~Martins, M.-F. Huang, T.~Okamoto, S.~Han, C.~Narisetty,
	J.~Fang, Y.-K. Huang, M.~Salemi \emph{et~al.}, ``Using global existing fiber
	networks for environmental sensing,'' \emph{Proceedings of the IEEE}, vol.
	110, no.~11, pp. 1853--1888, 2022.
	
	\bibitem{walter2020distributed}
	F.~Walter, D.~Gr{\"a}ff, F.~Lindner, P.~Paitz, M.~K{\"o}pfli, M.~Chmiel, and
	A.~Fichtner, ``Distributed acoustic sensing of microseismic sources and wave
	propagation in glaciated terrain,'' \emph{Nature communications}, vol.~11,
	no.~1, p. 2436, 2020.
	
	\bibitem{juarez2007field}
	J.~C. Juarez and H.~F. Taylor, ``Field test of a distributed fiber-optic
	intrusion sensor system for long perimeters,'' \emph{Applied optics},
	vol.~46, no.~11, pp. 1968--1971, 2007.
	
	\bibitem{hernandez2022deep}
	P.~D. Hern{\'a}ndez, J.~A. Ram{\'\i}rez, and M.~A. Soto, ``Deep-learning-based
	earthquake detection for fiber-optic distributed acoustic sensing,''
	\emph{Journal of Lightwave Technology}, vol.~40, no.~8, pp. 2639--2650, 2022.
	
	\bibitem{rorstadbotnen2023simultaneous}
	R.~A. R{\o}rstadbotnen, J.~Eidsvik, L.~Bouffaut, M.~Landr{\o}, J.~Potter,
	K.~Taweesintananon, S.~Johansen, F.~Storevik, J.~Jacobsen, O.~Schjelderup
	\emph{et~al.}, ``Simultaneous tracking of multiple whales using two
	fiber-optic cables in the arctic,'' \emph{Frontiers in Marine Science},
	vol.~10, p. 1130898, 2023.
	
	\bibitem{cedilnik2018advances}
	G.~Cedilnik, R.~Hunt, and G.~Lees, ``Advances in train and rail monitoring with
	das,'' in \emph{Optical Fiber Sensors}.\hskip 1em plus 0.5em minus
	0.4em\relax Optical Society of America, 2018, p. ThE35.
	
	\bibitem{milne2020analysis}
	D.~Milne, A.~Masoudi, E.~Ferro, G.~Watson, and L.~Le~Pen, ``An analysis of
	railway track behaviour based on distributed optical fibre acoustic
	sensing,'' \emph{Mechanical Systems and Signal Processing}, vol. 142, p.
	106769, 2020.
	
	\bibitem{thomas2023performance}
	P.~J. Thomas, Y.~Heggelund, I.~Klepsvik, J.~Cook, E.~Kolltveit, and T.~Vaa,
	``The performance of distributed acoustic sensing for tracking the movement
	of road vehicles,'' \emph{IEEE Transactions on Intelligent Transportation
		Systems}, pp. 1--14, 2023.
	
	\bibitem{wiesmeyr2020real}
	C.~Wiesmeyr, M.~Litzenberger, M.~Waser, A.~Papp, H.~Garn, G.~Neunteufel, and
	H.~D{\"o}ller, ``Real-time train tracking from distributed acoustic sensing
	data,'' \emph{Applied Sciences}, vol.~10, no.~2, p. 448, 2020.
	
	\bibitem{ferguson2020take}
	R.~J. Ferguson, M.~A. McDonald, and D.~J. Basto, ``Take the eh? train:
	Distributed acoustic sensing (das) of commuter trains in a canadian city,''
	\emph{Journal of Applied Geophysics}, vol. 183, p. 104201, 2020.
	
	\bibitem{tribaldos2021surface}
	V.~R. Tribaldos, J.~B. Ajo-Franklin, S.~Dou, N.~J. Lindsey, C.~Ulrich,
	M.~Robertson, B.~M. Freifeld, T.~Daley, I.~Monga, and C.~Tracy, ``Surface
	wave imaging using distributed acoustic sensing deployed on dark fiber:
	Moving beyond high-frequency noise,'' \emph{Distributed Acoustic Sensing in
		Geophysics: Methods and Applications}, pp. 197--212, 2021.
	
	\bibitem{bar1995multitarget}
	Y.~Bar-Shalom and X.-R. Li, \emph{Multitarget-multisensor tracking: principles
		and techniques}.\hskip 1em plus 0.5em minus 0.4em\relax YBs Storrs, CT, 1995,
	vol.~19.
	
	\bibitem{taweesintananon2021distributed}
	K.~Taweesintananon, M.~Landr{\o}, J.~K. Brenne, and A.~Haukanes, ``Distributed
	acoustic sensing for near-surface imaging using submarine telecommunication
	cable: A case study in the trondheimsfjord, norway,'' \emph{Geophysics},
	vol.~86, no.~5, pp. B303--B320, 2021.
	
	\bibitem{liu2017distributed}
	X.~Liu, C.~Wang, Y.~Shang, C.~Wang, W.~Zhao, G.~Peng, and H.~Wang,
	``Distributed acoustic sensing with michelson interferometer demodulation,''
	\emph{Photonic Sensors}, vol.~7, pp. 193--198, 2017.
	
	\bibitem{ester1996density}
	M.~Ester, H.-P. Kriegel, J.~Sander, X.~Xu \emph{et~al.}, ``A density-based
	algorithm for discovering clusters in large spatial databases with noise,''
	in \emph{kdd}, vol.~96, 1996, pp. 226--231.
	
	\bibitem{rockafellar2009variational}
	R.~T. Rockafellar and R.~J.-B. Wets, \emph{Variational analysis}.\hskip 1em
	plus 0.5em minus 0.4em\relax Springer Science \& Business Media, 2009, vol.
	317.
	
	\bibitem{sarkka2013bayesian}
	S.~S{\"a}rkk{\"a} and L.~Svensson, \emph{Bayesian filtering and
		smoothing}.\hskip 1em plus 0.5em minus 0.4em\relax Cambridge university
	press, 2023, vol.~17.
	
	\bibitem{vo2015multitarget}
	B.~N. Vo, M.~Mallick, Y.~Bar-Shalom, S.~Coraluppi, R.~Osborne, R.~Mahler, and
	B.~T. Vo, ``Multitarget tracking,'' \emph{Wiley encyclopedia of electrical
		and electronics engineering}, vol.~1, 2015.
	
	\bibitem{luo2021multiple}
	W.~Luo, J.~Xing, A.~Milan, X.~Zhang, W.~Liu, and T.-K. Kim, ``Multiple object
	tracking: A literature review,'' \emph{Artificial intelligence}, vol. 293, p.
	103448, 2021.
	
	\bibitem{burkard2012assignment}
	R.~Burkard, M.~Dell'Amico, and S.~Martello, \emph{Assignment problems: revised
		reprint}.\hskip 1em plus 0.5em minus 0.4em\relax SIAM, 2012.
	
	\bibitem{murphy2002dynamic}
	K.~P. Murphy \emph{et~al.}, ``Dynamic bayesian networks,'' \emph{Probabilistic
		Graphical Models, M. Jordan}, vol.~7, p. 431, 2002.
	
	\bibitem{mazor1998interacting}
	E.~Mazor, A.~Averbuch, Y.~Bar-Shalom, and J.~Dayan, ``Interacting multiple
	model methods in target tracking: a survey,'' \emph{IEEE Transactions on
		aerospace and electronic systems}, vol.~34, no.~1, pp. 103--123, 1998.
	
	\bibitem{johnston2001improvement}
	L.~A. Johnston and V.~Krishnamurthy, ``An improvement to the interacting
	multiple model (imm) algorithm,'' \emph{IEEE transactions on signal
		processing}, vol.~49, no.~12, pp. 2909--2923, 2001.
	
	\bibitem{challa2011fundamentals}
	S.~Challa, M.~E. Morelande, D.~Musicki, and R.~J. Evans, \emph{Fundamentals of
		object tracking}.\hskip 1em plus 0.5em minus 0.4em\relax Cambridge University
	Press, 2011.
	
\end{thebibliography}


\end{document}